\documentclass[a4paper,12pt]{article}

\usepackage{amsmath}
\usepackage{graphicx}
\usepackage{verbatim}
\usepackage[numbers]{natbib}
\usepackage{appendix}

\title{Ensembled Direct Multi-Step forecasting methodology with comparison on macroeconomic and financial data}
\author{Tomasz M. \L api\'nski \small{(84tomek@gmail.com)}\\S\&P Global Market Intelligence, \and Krzysztof Zi\'o\l kowski  \small{(krzysztof.ziolkowski@gdansk.merito.pl)}, \\WSB Merito University Gda\'nsk}

\begin{document}

\maketitle

\begin{abstract}
Accurate forecasts of macroeconomic and financial data, such as GDP, CPI, unemployment rates, and stock indices, are crucial for the success of countries, businesses, and investors, resulting in a constant demand for reliable forecasting models. This research introduces a novel methodology for time series forecasting that combines Ensemble technique with a Direct Multi-Step (DMS) forecasting procedure. This Ensembled Direct Multi-Step (EDMS) approach not only leverages the strengths of both techniques but also capitalizes on their synergy. The ensemble models in the methodology were selected based on performance, complexity, and computational resource requirements, encompassing a full spectrum of model complexities, from simple Linear and Polynomial Regression to medium-complexity ETS and complex LSTM model. Models were weighted based on their performances. In the DMS procedure we limit retraining to one- and five- year/month forecasts for economic and financial data respectively. Standard Iterative Multi-Step (IMS) procedure is employed for other horizons, effectively reducing computational demands while maintaining satisfactory results. The proposed methodology is benchmarked against Ensemble technique conventionally applied to IMS-generated forecasts, utilizing several publicly available macroeconomic and financial datasets. Results demonstrate a significant performance improvement with EDMS methodology, averaging a 33.32\% enhancement across the analysed datasets, and sometimes improvement reaching above 60\%.
\end{abstract}

\begin{keywords}
	Forecasting, time series, Direct Multi-Step, Ensemble, LSTM, ETS
\end{keywords}


\section{Introduction}
In recent years, time series forecasting has gained significant importance across various fields, particularly in economics, finance, and meteorology. This is primarily due to its crucial role in predicting future trends, which enables better decision-making and risk mitigation. For instance, accurate forecasts of key macroeconomic indicators, such as GDP, CPI, interest rates, and unemployment are essential for central banks, monetary authorities, and governments to guide monetary policy decisions, evaluate economic performance and labour market conditions, and plan government spending and taxation. Similarly, predicting future traffic flow or energy consumption facilitates improved planning in the economy, ensuring optimal resource allocation. Economic forecasting specifically, faces several challenges, including model uncertainty, non-stationarity, and the mixed frequency of data. In finance, forecasts are fundamental for establishing appropriate trading strategies, including hedging, speculation, and arbitrage. Private and corporate investors, businesses, commercial banks, and other financial institutions compute various forecasts daily, covering prices of stocks, indices, derivatives, and other financial assets. Lastly, in meteorology, forecasting weather conditions, including wind strength and weather anomalies is crucial for maritime transport and for protecting populations from unforeseen weather events.

Traditionally, forecasts have been computed as a one-step-ahead prediction problem, aiming to predict the value at the next time step, and longer-term forecasts are generated through the iterative application of the same model across consecutive time steps. This procedure is commonly referred as Iterative Multi-Step (IMS). The refinement of this procedure is when the model is retrained at each step by incorporating previously forecasted data, the approach is known as Direct Multi-Step (DMS).

Although DMS is less common than IMS, it offers significant advantages. Most importantly, DMS tends to enhance forecasts accuracy, enabling more precise macroeconomic forecasting and planning. According to \cite{chevillon_direct_multi_step} DMS is most beneficial when the model is misspecified, particularly when the data contains misspecified unit roots, neglected residual autocorrelation, and omitted location shifts. Retraining the model at each step in the forecast horizon generally makes the approach more specialized and tailored for specific use cases. Specifically, at each retraining step, the training set is extended with predictions from previous forecast executions, resulting in forecasts that are naturally more accurate than those generated by IMS. This improved accuracy is especially evident when dealing with non-stationary data or when model dynamics change over time \cite{marcellino_comparison_direct_iterated, mccracken_empirical_investigation_forecasts, chevillon_non_parametric_direct_multi_step}. However, the downside of this approach is the significant consumption of time and computing resources, as training is typically the most computationally demanding part of the forecasting process. Moreover, in DMS, we can utilize different models for each forecast step, which tend to perform better with longer forecast horizons \cite{chevillon_direct_multi_step, marcellino_comparison_direct_iterated, taieb_recursive_direct_multi_step}. Nonetheless, this approach presents additional challenges, such as reliance on varying sets of input parameters, model structures, or even learning algorithms, which adds complexity to the setup and execution of the training process \cite{taieb_recursive_direct_multi_step}.

In contrast, the traditional IMS method involves creating a model tailored to a specific date within the forecasting horizon and uses it repeatedly for consecutive time steps. This approach simplifies the forecasting process and, in some cases, can achieve similar accuracy to DMS, particularly when the data is stationary and the model is well-tuned \cite{chevillon_direct_multi_step, marcellino_comparison_direct_iterated}. Notably, iterated forecasting demonstrates better performance in terms of computing resources compared to DMS, as the model is trained only once. This is especially beneficial when dealing with numerous time series and long forecast horizons. However, as discussed in \cite{taieb_recursive_direct_multi_step}, the IMS approach can struggle to fit the data adequately, even when the model is perfectly specified for the training data. Moreover, IMS is more sensitive to model misspecification than DMS, as shown in \cite{taieb_recursive_direct_multi_step}, the Mean Squared Error (MSE) of IMS tends to be larger than that of DMS if the model is incorrectly specified.

DMS can be preferred over IMS, especially for macroeconomic forecasting, due to several practical and theoretical advantages. In IMS, the fact that forecasted value serves as an input for the subsequent forecast using the same model, leads to the accumulation of errors. A small inaccuracy in earlier predictions can accumulate over time, resulting in significantly less accurate forecasts for later periods. In contrast, DMS allows each model to be optimized for its target period, often leading to better performance, particularly in the context of macroeconomic indicators such as GDP, inflation, or employment. These indicators are influenced by numerous dynamic factors and frequently exhibit non-stationary behaviour. DMS forecasting can better accommodate these changes, as the model can be designed to incorporate recent trends and shifts in the data more dynamically. Furthermore, macroeconomic data often display nonlinear behaviour and are subject to structural breaks due to policy changes, economic shocks, or other external factors. DMS forecasting models are better equipped to handle these complexities because they do not rely on recursively generated inputs that may carry forward outdated or biased information.

A common technique for improving forecast accuracy is Ensembling. In this approach, an ensemble of models is utilized to compute forecasts rather than relying on a single model. The ensemble forecast is calculated as a weighted sum of the forecasts produced by individual models. The weights can either be equal or determined based on each model’s performance. Research indicates that ensembled forecasts tends to generalize better to unseen data, as this technique leverages the strengths of multiple models while complementing their weaknesses. Consequently, the ensembled forecast is typically more accurate than those generated by individual models.

We propose a new forecasting methodology called Ensembled Direct Multi-Step (EDMS), which combines Direct Multi-Step approach with Ensemble forecasting technique. Similar to DMS method, EDMS involves retraining models. However, we limit retraining to specified time steps, and also for this time steps forecasts are generated using the Ensemble technique. For the remaining time steps, forecasts are iteratively computed, as in IMS method. In general, retraining in EDMS can be performed at arbitrary time steps throughout the forecasting horizon. Reducing the number of retrainings is particularly important when forecasts consume significant computing resources. This situation often arises when dealing with large training datasets, such as an extensive number of time series or observations, which is common in economic and financial datasets. The models included in the ensemble are carefully selected based on criteria such as accuracy, complexity, and computational efficiency. When ensembling, each model contributes to the ensemble according to its weight, which reflects its performance. This ensures that the most accurate forecasts have the greatest impact on the ensemble. To assess the practical viability of EDMS, we conduct an extensive comparative analysis of this methodology against the conventional Ensemble technique that utilizes IMS (EIMS). This analysis is conducted on common and publicly available macroeconomic and financial datasets. 

The remainder of the paper is organized into several sections: Section 2 offers a comprehensive literature review of the methodology's components, specifically Ensembled forecasting method, models utilized in the ensemble, and IMS and DMS procedures, as well as some improvements of DMS. Section 3 details the EDMS methodology, including a description of the models employed in the ensemble and the Ensemble technique itself. In Section 4, the document presents a comparative analysis of macroeconomic data, evaluating the performance of the EDMS methodology in relation to the conventional Iterative Multi-Step forecasting approach. This section also characterizes the datasets used and the comparison methodology. At the end of Section 4, the results are presented, showing significant improvements in forecasting accuracy with the use of the EDMS approach. The final, fifth section, concludes with a summary of the key findings of the study, as well as potential improvements, extensions and applications.

\section{Literature review}

In the literature, there is a wide variety of techniques available for forecasting time series. Recently, much of the research focus is on machine learning (ML) models, particularly Artificial Neural Networks (ANNs). While traditional models, such as ARIMA and ETS, have been successfully used for years, the application of ANNs offers significant improvements in many aspects of forecasting, including capturing complex temporal patterns. Notably, the most researched ANN model is Long Short-Term Memory (LSTM), which has been well-established to produce accurate forecasts \cite{park_forecasting_topic_trends, li_enhancing_financial_forecasting, jiang_trend_pattern_fuzzy_lstm, shi_cnn_lstm_carbon_price_forecasting, rahman_integrating_deep_learning_algorithms, chen_time_series_forecasting_oil_production, sushmi_real_time_irradiance_forecasting, li_time_series_production_forecasting}. Furthermore, LSTM is often utilized as components of hybrid approaches, such as in combination with ARIMA. Several articles demonstrate that such hybrid approaches perform better than the individual models alone [38,39,40,41].

Furthermore, when it comes to forecasting economic time series, machine learning models frequently outperform traditional econometric models, such as VAR or (S)ARIMA(X). In particular, ANN architectures, especially LSTM, has proven to be well-suited for this purpose \cite{taieb_review_comparison_strategies, ahmed_empirical_comparison_ml_models, de_gooijer_25_years_forecasting, li_recent_developments_econometric_modeling}. Additionally, the literature presents numerous comparisons of a wide range of machine learning models, including XGBoost, LSTM, GRU, and DeepAR, against traditional ARIMA. In almost all cases, ML models demonstrate superior performance compared to the traditional counterpart \cite{aziz_facebook_prophet, gumus_crude_oil_forecasting, kashpruk_comparative_research_statistical_models, yun_interpretable_stock_price_forecasting,kumar_prediction_area_production}. One explanation for this phenomenon is variety of available models, so that the model structure and architecture can be tailor to specific use case. Moreover, high dependence of machine learning on hyperparameters selection allows for extensive tuning and offers the potential to significantly enhance accuracy.

Numerous comparisons have been made between the well-known Prophet model \cite{prophet_github} and both, ML and traditional econometric models. It is important to note that Prophet is an enhancement of traditional models. Specifically, Prophet generalizes the additive model, where different characteristics of the time series, such as trend, seasonality, holidays, and special events, are combined, along with the addition of random noise \cite{prophet_github}. Some studies assert that Prophet outperforms traditional models \cite{cheng_forecasting_bitcoin_prices, ghosh_clean_energy_stock_price_forecasting, albahli_lstm_vs_prophet, kwarteng_comparative_analysis_arima_sarima_prophet, brykin_sales_forecasting_models, sharma_time_series_forecasting_fb_prophet}, while others suggest that ML or hybrid ML models are superior \cite{aziz_facebook_prophet, gumus_crude_oil_forecasting, kashpruk_comparative_research_statistical_models, yun_interpretable_stock_price_forecasting, kumar_prediction_area_production, guruge_time_series_forecasting_kubernetes}. The reason behind Prophet's strong performance lies in its specific design, which is also particularly well-suited for handling seasonal data effectively.

Ensemble forecasting is a widely adopted strategy for forecasting. This technique aggregates predictions from multiple models. With such a approach forecasts consistently outperform individual models forecasts in terms of accuracy and stability across diverse domains. The seminal work of \cite{bates_combination_1969} first demonstrated the benefits of combining forecasts, highlighting significant improvements in predictive performance. Subsequent studies by \cite{newbold_experience_1974, shi_1999} confirmed the effectiveness of both linear and nonlinear combinations. Building on this foundation, \cite{kourentzes_improving_2014a, kourentzes_neural_2014b} showed that combining models trained on data sampled at different frequencies enhances forecasting accuracy across short-, medium-, and long-term horizons, with minimal differences between mean and median operators. Additional empirical evidence from studies such as \cite{weng_macroeconomic_2018, booth_automated_2014, qian_stock_2007, chen_flexible_2007, gocken_integrating_2016, patel_predicting_2015, kristjanpoller_volatility_2014, azqueta_news_2017, saltzman_uncertainty_2018, tobback_belgian_2016, ensemble_1, ensemble_2, ensemble_3, ensemble_4} and \cite{time_series_analysis_1}, p. 145, further shows the effectiveness of ensemble methods in improving forecast performance.

The concept of Direct Multi-step forecasting procedure has been around for a relatively long time, and the debate on IMS and DMS remains active, with various studies comparing their performance. Many research studies indicate that DMS outperforms IMS, however, there are also examples demonstrating the opposite.
According to the work of Chevillion and Hendry \cite{chevillon_non_parametric_direct_multi_step}, early research conducted by Klein \cite{klein_essay_theory_economics_prediction} and Johnston, Klein, and Shinjo \cite{johnston_estimation_prediction_dynamic_models} suggests that DMS might be more efficient than IMS. These results were inspired by the application of Cox's \cite{cox_prediction_exponentially_weighted} ideas to exponentially-weighted moving-average (EWMA) or integrated moving-average (IMA (1,1)) models \cite{chevillon_non_parametric_direct_multi_step}. In extensive empirical studies, Kang \cite{kang_multiperiod_forecasting} and Marcellino et al. \cite{marcellino_comparison_direct_iterated_micro} observed only minimal improvements with DMS, while Proietti \cite{proietti_direct_iterated_multistep} extended these findings to other models, i.e. ARIMA, and reached a similar conclusion. According to works of Bhansali \cite{bhansali_direct_autoregressive_predictors} and Schorfheide \cite{schorfheide_forecasting_under_misspecification}, under certain assumptions, DMS can outperform IMS. Bhansali asserts that for a finite autoregressive process, IMS is superior; however, for other models, DMS can effectively reduce the mean squared error of prediction when using an under-parametrized model. In contrast, Schorfheide demonstrates that DMS forecasting yields favourable results for series related to wages, prices, and money. Furthermore, Chevillon (2017) highlights that DMS models are particularly effective when managing the bias-variance trade-off over extended forecasting horizons. In contrast, more recently, McCracken et al. \cite{mccracken_empirical_investigation_forecasts} compared the accuracy of direct multi-step and iterated multi-step forecasting using VAR and ARDL models, specifically VAR-based IMS and ARDL-based DMS. They concluded that IMS approaches generally perform better, however, DMS-based methods show promise for forecasting hypothetical scenarios, such as the influence of oil prices on inflation in a VAR model and for nominal variables.

Numerous studies indicate that DMS tends to yield better results when the employed model misrepresents the underlying data-generating process of the modelled phenomena. Moreover, Lin and Tsay \cite{lin_tsay_co_integration_constraint} demonstrated that certain types of model misspecification can lead to improved forecasts when using DMS instead of IMS \cite{tsay_comment_adaptive_forecasting, kang_multiperiod_forecasting, marcellino_comparison_direct_iterated_micro, chevillon_non_parametric_direct_multi_step}. A similar conclusion was reached in Tiao and Xu's \cite{tiao_robustness_maximum_likelihood} paper, which states that DMS is more efficient than IMS when the model is misspecified. Conversely, when the model is correctly specified, Johnston \cite{johnston_estimation_prediction_dynamic_models} indicated that IMS performs better \cite{cox_prediction_exponentially_weighted, klein_essay_theory_economics_prediction, johnston_estimation_prediction_dynamic_models, chevillon_non_parametric_direct_multi_step}. In contrast, Weiss \cite{weiss_multi_step_estimation} explored DMS and ARIMA models, finding that when the model is incorrectly specified, both IMS and DMS yield relatively good results. However, in small samples, when the number of observations is limited, DMS tends to perform worse \cite{findley_use_multiple_models, weiss_multi_step_estimation, weiss_andersen_estimating_time_series, chevillon_non_parametric_direct_multi_step}. Further insights regarding misspecification are provided by Chevillon \cite{chevillon_direct_multi_step}, who concludes that DMS is a valuable technique for forecasting, particularly in the presence of model misspecification and non-stationarity. He also found that the advantages of IMS or DMS vary depending on the forecast horizon and the stochastic properties of the data. When comparing DMS to IMS, McElroy \cite{mcelroy_direct_iterative_identical} argues that the differences in performance between direct and iterative forecasting primarily stem from the model fitting method and the subtleties of model misspecification. Therefore, the choice of forecasting method should be made with careful consideration of the model's specifics and the context of its application.

Additionally, literature provide us with some improvements of DMS procedure. For instance, In and Jung \cite{in_jung_simple_averaging} propose a mixed approach of DMS with IMS, known as the Direct and Recursive Forecast Averaging Method (DRFAM). This method integrates various strategies for forecasting daily product sales over the next 28 days using hierarchical data. The features of both DMS and IMS are essential to the DRFAM model, and their combined application improves forecasting accuracy when addressing challenges associated with hierarchical data. Furthermore, Dossani \cite{dossani_inference_multi_step_forecasting} introduces a novel method for inference in direct multi-step forecasting regressions. This approach involves estimating the covariance matrix in direct multi-step and long-horizon forecasting regressions, which enhances both inference accuracy and efficiency. The proposed estimator significantly improves efficiency and accuracy compared to conventional HAC methods (heteroskedasticity and autocorrelation consistent covariance estimator) for both direct multi-step and long-horizon forecasts. The inference method presented in that paper enhances forecasting by providing more accurate estimates of uncertainty, i.e. standard errors, around forecasted values, thereby increasing confidence in predictive models. Clark and West \cite{clark_west_approx_normal_tests} investigate the performance of DMS forecasting in nested models, with a particular focus on utilizing mean squared prediction error (MSPE). Their paper compares two nested linear forecasting models: a parsimonious model with fewer parameters and a model that includes additional predictors. The authors evaluate the out-of-sample predictive performance of these models using adjusted mean squared prediction error (MSPE). The final empirical applications demonstrate that MSPE-adjusted measure provides a reliable assessment of forecast accuracy in nested models, underscoring its practical relevance in forecasting. Econometricians continue to explore both, direct and iterated, multi-step forecasting methods, with significant contributions from Findley, Weiss, Tiao, Xu, Lin, Granger, Clements, Hendry, Bhansali, Kang, Chevillon, and Schorfheide.

\section{Ensembled Direct Multi-Step forecasting}	\indent  In this section, we describe in detail Ensembled Direct Multi-Step (EDMS) forecasting method. Subsection 2.1 presents models used for the ensemble forecasts, Subsection 2.2 discusses Ensembling method itself, and finally, Subsection 2.3 explores application of Direct Multi-Step (DMS) forecasting combined with Ensembling method.
	\subsection{Models}
\begin{figure}[htbp]
			\centering
			\includegraphics[width=0.8\textwidth]{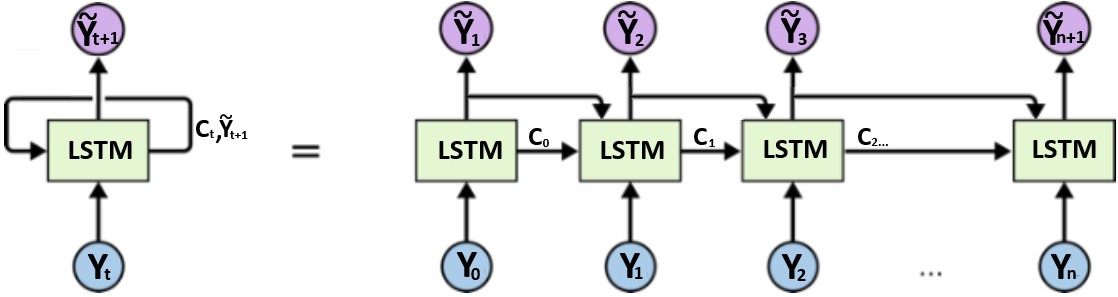}
			\caption{LSTM recurrent architecture, adapted from \cite{lstm_images} licensed under CC BY 4.0.}
			\label{fig:lstm_architecture}
		\end{figure}

\begin{figure}[htbp]
			\centering
			\includegraphics[width=0.8\textwidth]{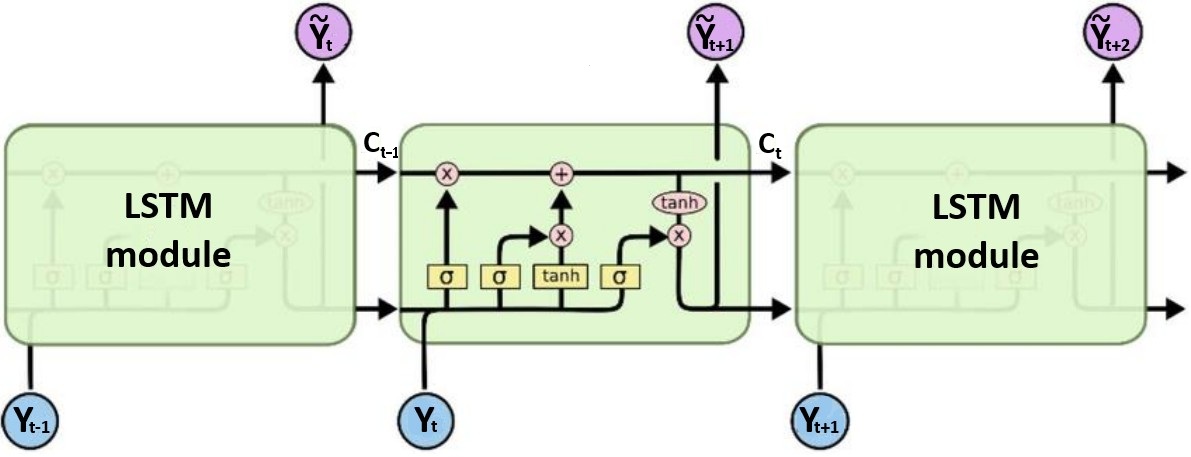}
			\caption{LSTM repeating module structure, adapted from \cite{lstm_images} licensed under CC BY 4.0.}
			\label{fig:lstm_module}
		\end{figure}

		 To enhance the modelling capability of the ensemble, we carefully selected a diverse set of models. We considered only models that perform well in terms of accuracy and computational efficiency. Additionally, we included models across full spectrum of complexity, ranging from simple to highly complex. For the simplest models, we employed basic forecasting using the average of past growth rates, as well as Linear and Polynomial Regression. For models of medium complexity, we utilized classical time series models, specifically General Exponential Smoothing (ETS). For the most complex model, we implemented an deep learning model, namely LSTM. In selecting these models, we aimed to minimize the computing resources and time required for forecasting, as economic and financial datasets tend to be relatively large. At the same time, we strive to maintain the highest possible accuracy. Prophet was excluded from the set of models due to significant computation time and gave minimal, if any, improvement in accuracy. We also have a constraint in the model selection that the datasets must be sufficiently large, as a considerable amount of data is necessary to train more complex models such as LSTM.\\
		\indent\quad For every model we compute two forecasts:
		\begin{enumerate}
			\item Performance forecast - these forecasts are calculated to assess the performance of the methods employed for computing weights in the ensembling process. We allocate 80\% of the available data for training and 20\% for testing. This split ratio was selected to ensure that there is sufficient data for effective model training, while also maintaining a reasonably large test set for proper model comparison. Experiments have demonstrated that the 80/20 ratio yields favourable results. Once forecasts are computed using the model trained on the training data, the value of performance metric is computed based on testing data
			\item Full forecast - this forecasts are used to compute the ensembled forecast. We utilize 100\% of the data for training and compute desired forecasts with the model trained on the complete dataset in order to get the best possible forecast.
		\end{enumerate}
		Let us now individually describe each model that is used in the ensemble:
		\paragraph{LSTM}
		This is a type of Recurrent Neural Network (RNN) model, which is a natural choice for modelling time series due to its ability to effectively recognize and learn patterns in sequential data. Notably, LSTM is well-known and widely used model for this purpose, as outlined in \cite{lstm_description}. We chose this model for several reasons. First, according to a literature review in \cite{systematic_review}, LSTM is the most extensively studied deep learning model for time series forecasting, and there is a lot of information to support it's usage. Particular reason for this selection is performance. Multiple sources indicate that LSTM outperforms widely used classical model ARIMA, as noted in \cite{lstm_arima_comparison_1, lstm_arima_comparison_2, lstm_arima_comparison_3, lstm_arima_comparison_4}. Furthermore, LSTM often prove to be the best performers among deep learning models, particularly for economic and financial data, as seen in \cite{lstm_arima_comparison_4, lstm_ml_comparison}.\\
		\indent\qquad The Long Short-Term Memory ANN architecture consists of sequentially connected modules with the number $n$ of these modules equal to the size of the input sequence $(y_{1}, y_{2}, \ldots, y_{n})$.  In this sequence, each module takes the following inputs:
		\begin{itemize} 
			\item Input element $y_{t}$ - the element of the input sequence at time $t$,
			\item Previous output $\widetilde{y}_{t}$ - the output of the previous module, which is the previous forecast at $t$,
			\item Previous cell state $C_{t-1}$ - the cell state of the previous module, which facilitates long term memory. 
		\end{itemize}
		See Figure \ref{fig:lstm_architecture} for illustration of LSTM model architecture.\\
		A single module is composed of the following components:
		\begin{enumerate}
			\item Forget gate - a fully connected layer with sigmoid activation $\sigma$, defined as:
				\begin{equation*}
					\sigma(y) = \frac{1}{1 + \exp(-y)},
				\end{equation*}
				where $y$ is the input signal. This sigmoid layer determines which parts of the cell state to forget, using the previous output $\widetilde{y}_{t}$ and the input sequence element $y_{t}$,
				\begin{equation*}
					f_{t}=\sigma(y_{t}U_{f}+\widetilde{y}_{t}W_{f})
				\end{equation*}
				where $U_{f}, W_{f}$ are weight matrices with weights parameters which determined in the training. This gate is depicted as first from left yellow rectangular in the module in Figure \ref{fig:lstm_module}.   
			\item Update gate - this gate first computes a fully connected layer with hyperbolic tangent activation $tanh$ that proposes information $\widetilde{C}_{t}$ for the new cell state, where 
				\begin{equation*}
					\tanh(y) =  \frac{\exp(y) - \exp(-y)}{\exp(y) + \exp(-y)}. 
				\end{equation*}
				Update gate also computes a fully connected layer with sigmoid activation $\sigma$, known as the input layer $i_{t}$, selects which of the proposed information to add to the cell state. Both layers use previous output $\widetilde{y}_{t}$ and the element of the input sequence $y_{t}$ as inputs. Then this gate combines these two outputs to compute an update to the state $C_{t}$. Those computations can be expressed as follows:
				\begin{align*}
					\widetilde{C}_{t}&=\tanh(y_{t}U_{C}+\widetilde{y}_{t}W_{C}),\\
					i_{t}&=\sigma(y_{t}U_{i}+\widetilde{y}_{t}W_{i}),\\
					C_{t}&=i_{t}\widetilde{C}_{t}+f_{t}C_{t-1},\\
				\end{align*}
				where $U_{i}, W_{i}$ and $U_{C}, W_{C}$ are weight matrices, with weights parameters which determined in the training. This gate is depicted as two connected middle yellow rectangulars together with connection with Input Gate in the module in Figure \ref{fig:lstm_module}.   
			\item Output gate - this final gate consists of a fully connected layer with sigmoid activation $\sigma$, with $\widetilde{y}_{t}$ and $y_{t}$ as inputs, which selects which parts of the cell state $C_{t}$, processed through $\tanh$ layer, to output :
				\begin{align*}
					o_{t}&=\sigma(y_{t}U_{o}+\widetilde{y}_{t}W_{o}),\\
					\widetilde{y}_{t+1}&=o_{t} \tanh(C_{t})
				\end{align*}
				where $U_{o}, W_{o}$ are weight matrices with weights parameters which determined in the training. This gate is depicted as first from the right yellow rectangular in the module, together with connection to the output, in Figure \ref{fig:lstm_module}.   
		\end{enumerate}
		The size and dimensions of $y_{t}, \widetilde{y}_{t}$ are determined by the specific case at hand, which also determines the dimensions of $C_{t}, \widetilde{C}_{t}, i_{t}, o_{t}$ and the size of weight matrices. For a detailed description, see \cite{lstm_description}. Refer to Figure \ref{fig:lstm_module} for illustration of LSTM module.
		\\
		\paragraph{ETS}
		This model is known as triple exponential smoothing, where ETS stands for Error, Trend, and Seasonality terms. This is generalization the Holt-Winters model by incorporating multiplicative error and trend terms. The term "Exponential Smoothing" derives from the fact that the model's equations can be expanded to represent an average of past observations with exponentially decaying weights. We chose this model for its reliability and widespread use, offering an excellent balance between performance and computational efficiency. For a detailed description and justification of this choice, see \cite{time_series_analysis_1}. For short series, specifically those with fewer than 50 observations, ETS and its special case, Holt's model, may be the most effective, as this quantity of data is typically sufficient to train models with reasonable accuracy, further details can be found in \cite{time_series_analysis_1, time_series_analysis_2}. When 50 or more observations are available, more complex models may be necessary, we address that by using LSTM. In our experiments, we selected Holt's model, which is additive version of ETS without seasonality, as it yielded the best results and has been proven very effective in term of computation time. We excluded seasonality because the series in our experiments generally lack seasonal variation. Let us briefly describe Holt's model:

		\begin{align*}
			L_{t}&=\alpha y_{t} + (1-\alpha)(L_{t-1} + T_{t-1}),\\
			T_{t}&=\gamma(L_{t} - L_{t-1}) + (1 - \gamma) T_{t-1},
		\end{align*}
		where $L_{t}, T_{t}$ represent the Level and Trend terms, respectively, and $\alpha, \gamma$ are smoothing parameters that typically fall within the range $(0,1)$. The prediction $h$ time steps ahead is given by:
		\begin{equation*}
			\widetilde{y}_{t+h}=(L_{t} + hT_{t}).
		\end{equation*}
		The smoothing parameters can be computed using standard methods, such as minimizing RSS for historical data. The starting values $L_{1}$ and $T_{1}$ are computed using first few observations in the series, e.g. $L_{1}=\sum_{i=1}^{s}x_{i}/2$.
		\paragraph{Linear and Polynomial Regression}
	         Linear Regression is one of the most widely used and most established statistical models. So we considered this model as a candidate for a model of low complexity. As expected, model yielded a decent results for simple and regular time series. Polynomial Regression is analogous to Linear Regression but fits the data with a quadratic function instead of a linear one. Similar to Linear Regression, this model is straightforward and yields good results for rather regular time series, and was particularly good for short prediction horizons.
		\paragraph{Average Growth Rate}
	          A simplistic method is used to enrich the ensemble. The method computes the historical growth rate on the training series using taking median of the relative changes of values from one observation to other, that is, the following expression:
		\begin{equation*}
					\delta(t)=1+\frac{y_{t}-y_{t-1}}{y_{t-1}}
		\end{equation*}
		where $t$ is time of single observation. So the historical growth rate is a median of values $\widehat\delta = \text{median}(\delta(t), t=1,\ldots, N)$. Predictions are generated by multiplying the last observation by the average growth rate and repeating this for successive forecast dates:
		\begin{equation*}
					\widetilde{y}_{t+h}=y_{t}\widehat\delta^{h}
		\end{equation*}
		where $h$ is the number of observation in the forecast horizon.
	\subsection{Ensembling}
		\label{section_ensembling}
\begin{figure}[htbp]
			\centering
			\includegraphics[width=0.8\textwidth]{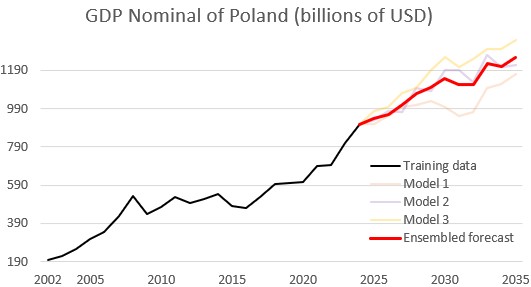}
			\caption{Ensemble forecast from predictions of three models}
			\label{fig:Ensembling_illustration}
		\end{figure}

\begin{figure}[htbp]
			\centering
			\includegraphics[width=0.8\textwidth]{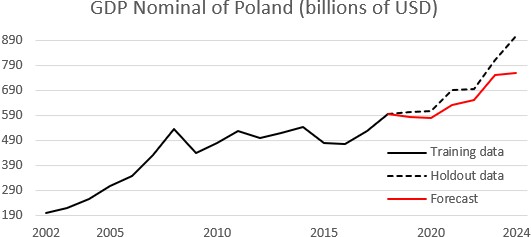}
			\caption{Forecast for testing dates}
			\label{fig:Ensemble_weight}
		\end{figure}

\begin{figure}[htbp]
			\centering
			\includegraphics[width=0.8\textwidth]{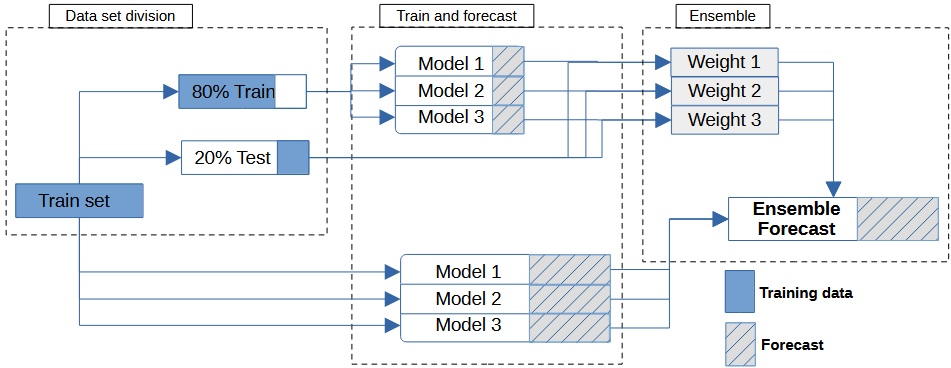}
			\caption{Diagram of Ensemble technique}
			\label{fig:Ensemble_technique_diagram}
		\end{figure}

			As mentioned in the literature review, numerous articles demonstrates that ensembles are often more accurate than any single model on its own. The process of pooling models for the ensemble involves computing accuracy scores for each model. These models are then combined into a single prediction using a weighted arithmetic mean, as described by the formula:
		\begin{equation}
			\label{ensembling}
			\widehat{y}_{t+h}=\sum_{i}w_{i}\widetilde{y}_{t+h, i},
		\end{equation}
		where $w_{i}$ represents the weights and $\widehat{y}_{t+h,i}$ denotes the forecasts at time step $t$ and $h$ steps ahead for the $i$-th model.\\
		Refer to Figure \ref{fig:Ensembling_illustration} for an illustration of the ensemble technique.\\
		Since the weights of the models are computed based on accuracy scores, we can ensure more accurate models receive a higher weight. The metric employed in computation of the score is the Mean Absolute Error (MAE). We also experimented with other metrics, including Mean Squared Error, Root Mean Squared Error, and Mean Absolute Percentage Error. However, the MAE metric yielded the best performance in the final results. \\
		\indent Since higher metric values indicate greater errors, we need to invert these values so that a higher metric value reduces the model's weight:
		\begin{equation*}
			\label{performance_mape_inversion}
			mae'_{i}= 1-\frac{mae_{i}}{\sum_{i}mae_{i}},
		\end{equation*}
		and so we have the following expressions for the weights:
		\begin{equation}
			\label{performance_weights}
			w_{i}=\frac{mae'_{i}}{\sum_{i}mae'_{i}},
		\end{equation}
		where $mae_{i}, mae'_{i}, w_{i}$ represent the values of the metric, inverted metric, and weight for the $i$-th model, respectively. To compute the metric value, we separate 20\% of the available dataset into test data. The model is trained on the dataset reduced to 80\%, and forecasts are generated for the same dates as the testing data. The errors of the predictions relative to the testing data are computed using the metric, as expressed in the following expression:
		\begin{equation}
			\label{performance_metric}
			mae_{i}=\frac{1}{H}\sum_{h=1}^{H} \big|\widehat{y}_{t+h,i}-y_{t+h}\big|,
		\end{equation}
		where $H$ is the number of observations in the forecasting horizon, $\widehat{y}_{i,t+h}$ are the forecasts at time step $t$ and $h$ steps ahead for the $i$-th model, $y_{t+h}$ is testing data at time step $t$ and $h$ steps ahead. Refer to Figure \ref{fig:Ensemble_weight} for exemplary testing forecasts alongside the corresponding testing data.\\
		See Figure \ref{fig:Ensemble_technique_diagram} for a diagram illustrating how the ensemble technique is carried out.
		The success of the ensemble technique lies in the fact that models in the ensemble tends to complement each other. Forecasts of well-performing models are amplified by assigning them greater weights, thereby exerting a more significant influence on the combined forecast. However, even high-quality forecasts have their flaws. In certain data segments, other models, despite having lower overall performance, may demonstrate more desirable behaviour. Thus, by averaging, these less-performing models enhance accuracy with a relatively minor effect on the strengths of the well-performing models, as they receive lower weights. In summary, through the process of averaging, this method has the capability to amplify the strengths, and with minimal impact to overall improvement, also mitigate the weaknesses, of individual models.
	\subsection{Direct Multi-Step (DMS)}
\begin{figure}[h!tbp]
			\centering
			\includegraphics[width=0.8\textwidth]{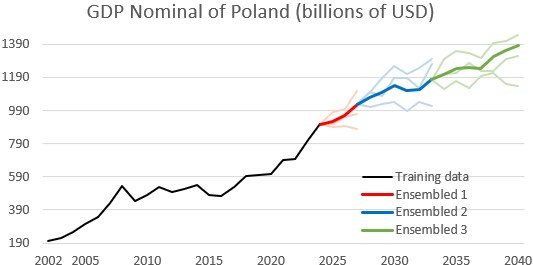}
			\caption{Exemplary illustration of Ensembled DMS}
			\label{fig:EDMS}
		\end{figure}

\begin{figure}[h!tbp]
			\centering
			\includegraphics[width=0.8\textwidth]{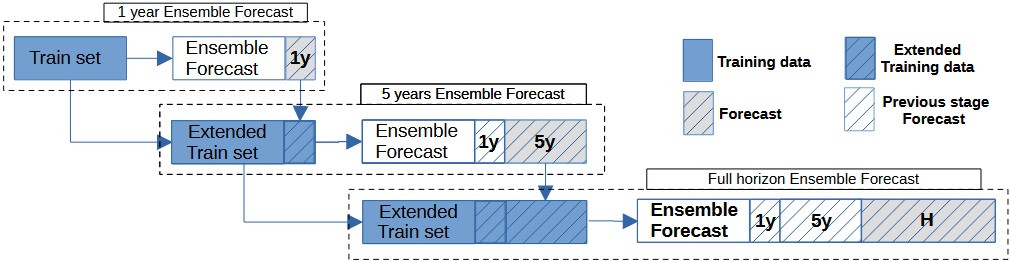}
			\caption{Diagram of EDMS methodology. Ensemble forecasts as depicted in Figure \ref{fig:Ensemble_technique_diagram}}
			\label{fig:EDMS_diagram}
		\end{figure}

		In a typical DMS procedure, a forecast is first computed one step ahead. The resulting forecast is then added to the training data, and the model is retrained on this extended dataset. This procedure is repeated for the number of observations in the forecasting horizon. Retraining is the most computationally expensive part of forecast, hence, to decrease the amount of computing resources used, we refrain from applying DMS at each step. Specifically, we retrain the models at two key intervals for economic and financial data respectively: once after the first year/month to generate 5-year/month forecasts, and again after five years/month to produce 20 to 30-year/month forecasts. These horizons are can be seen as commonly recognized short and medium-term economic and financial forecasting periods. For the remaining steps, we employ the standard IMS procedure, which simply shifts the forecast input by one observation without retraining the model or using ensembling. This limited retraining schedule allows us to retain the benefits of DMS, including its robustness to non-stationary data and structural changes, while significantly reducing the computational burden typically associated with full DMS implementations. This methodology can be decomposed into the following three-step algorithm, where, for illustrative purposes, we specify the frequency to be annual with retraining at the first and fifth years:
	\begin{enumerate}
		\item Step 1 - One year ahead forecast - these forecast are computed for each model in the ensemble. The ensemble technique described in subsection \ref{section_ensembling} is then applied. Specifically, the available data is divided into training and testing sets in an 80/20 ratio, and forecasts are computed for the testing observations. The MAE metric is used to compute performance weights according to formulas (\ref{performance_weights}) and (\ref{performance_metric}). Finally, the one-year forecasts are ensembled using the calculated weights, resulting in the ensembled forecast according to (\ref{ensembling}). 
			\begin{equation*}
				\label{ensembling_1year}
				\widehat{y}_{t+1}=\sum_{i}w_{1,i}\widetilde{y}_{t+1, i}(Y_{t}),
			\end{equation*}
			where $w_{i,1}$ are computed weights and $\widetilde{y}_{t+1,i}$ are the forecasts at the time step one year ahead from time $t$ for the $i$-th model, which is trained on the training data up to time $t$, denoted as $Y_{t}=(y_{1}, y_{2}, \ldots, y_{t})$.
		\item Step 2 - Retrain models and forecast five year ahead - the models are retrained with the training data extended by the ensembled forecast from Step 1. Then forecasts up to five-year-ahead horizon are then computed using the IMS procedure and ensembled in the same manner as in Step 1. The explicit formula for the forecast is given by the following expression:
			\begin{equation*}
				\label{ensembling_5year}
				\widehat{y}_{t+1+h}=\sum_{i}w_{5,i}\widetilde{y}_{t+1+h, i}(Y_{t},\widehat{y}_{t+1}),
			\end{equation*}
			where $h=1,2,3,4$ are the next time steps in the forecasting horizon, $w_{i,5}$ are computed weights for five year forecasts, and $\widetilde{y}_{t+1+h,i}$ are the forecasts at the time steps up five years ahead from time $t+1$, for the $i$-th model, trained on the initial training data $Y_{t}$ and the forecast from previous step.
		\item Step 3 - Retrain and forecast full horizon - the model is retrained on the training data further extended by the ensembled forecasts from Step 2. Ensembled forecasts up to the full horizon are then computed as previously described. The expression for the forecasts is as follows:
			\begin{equation*}
				\label{ensembling_Hyear}
				\widehat{y}_{t+5+h}=\sum_{i}w_{H,i}\widetilde{y}_{t+5+h, i}(Y_{t},\widehat{Y}_{t+5}),
			\end{equation*}
			where $h=1,2, \ldots, H-5 $ are the next time steps in the forecasting horizon, $w_{H,i}$ are the computed weights for full horizon, and $\widetilde{y}_{t+5+h,i}$  are the forecasts for each model, trained on data further extended by forecasts up to the fifth year, denoted as $\widehat{Y}_{t+5}=(\widehat{y}_{t+1},\widehat{y}_{t+2},\ldots,\widehat{y}_{t+5})$.
	\end{enumerate}
	The above algorithm can be easily adapted to other data frequencies, such as quarterly, monthly, and daily, by appropriately increasing the number of data points. The retraining periods can also be adjusted, for example, in our experiments with financial data, we chose one- and five-month intervals instead of one- and five-year forecasts. See Figure \ref{fig:EDMS} for an illustration of the methodology in plots and
	Figure \ref{fig:EDMS_diagram} for illustration of EDMS methodology on diagram.
	\\
	\indent Typical DMS, as highlighted in the literature review, generally enhance performance, particularly over medium- and long-term horizons. In contrast IMS, even when combined with ensemble forecasting, tend to accumulate errors. When initial forecasts contain relatively small errors, feeding these forecasts back into the same model to generate subsequent predictions can amplify those initial inaccuracies, as the model retains its inherent flaws. Given the substantial body of research on DMS, it is evident that this procedure possess the capability to mitigate this accumulation of errors, with the reduction effect becoming increasingly significant as the forecasting horizon extends. This effect is especially visible in macroeconomic forecasting, encompassing time series such as GDP, inflation, and employment over long horizons. Macroeconomic time series often exhibit non-linear and non-stationary behaviour and are subject to structural breaks resulting from policy changes, economic shocks, or other external influences. DMS forecasting is better suited to handle these complexities because it do not depend on recursively generated inputs that may carry forward outdated or biased information. Instead, the procedure is designed to directly incorporate recent trends and shifts in the data.\\
	\indent Moreover, applying this EDMS methodology should create a synergistic effect on performance improvement, surpassing the advantages of each method when used independently. The Ensemble technique, which employs performance-based weighting of models, allows the ensemble to dynamically emphasize models that perform better across different horizons and data characteristics. Specifically, at each retraining step, the weights are recalculated based on the extended data, making them more aligned with the forecasts currently being executed, thereby enhancing the impact of the traditional Ensembling technique. Consequently, each DMS retraining is conducted on a dataset extended with more accurate data than what is typically used in DMS, as the forecasts have been improved through ensembling. As a result, the trained model is more closely tailored to the underlying phenomena that generates data, than a standard DMS model. 
\section{Comparisons on macroeconomic and financial data}
	We evaluate EDMS methodology against the EIMS approach to quantify the performance gains achieved through direct forecasting. In Subsection 4.1, we describe the data used, along with the process of data gathering and preparation. Subsection 4.2 explains the methodology employed in the comparative analysis, while the final Subsection 4.3 presents and discusses the results.
	\subsection{Data}
		 Our comparative analysis utilizes carefully selected, common, and publicly available datasets. We chose public data for transparency and to ensure that our methodology is tested with data that is relatively similar to data potentially used in the practical applications. Moreover, to ensure the objectivity of our results, we employ a well-defined process of data selection and preparation, which can be broken down into the following stages: 
		\begin{enumerate}
			\item Indicators selection - a set of common macroeconomic indicators is selected.
			\item Source selection - for the selected indicators, we identify the public source with the highest quality of data, specifically focusing on methodologically consistent datasets across multiple countries for the same indicators. Furthermore, countries are selected based on two primary criteria: geographical diversity and the level of economic development, aiming to capture a representative cross-section of the global economy.
			\item Trimming data set - for the forecasting implementation, it is essential that all time series have equal lengths. However, the time series in the dataset vary in length. We refrain from reducing all time series to the shortest one, as this would significantly decrease the size of the dataset and ultimately diminish the quality of model training. Instead, we remove only those time series with the smallest number of observations while simultaneously reducing the overall length of all time series in a manner that maximizes the number of data points in the entire dataset. 
			\item Exclude outliers - we eliminate time series that are highly irregular and cause model fit errors that are disproportionately higher than those of other time series in the specific dataset.
		\end{enumerate}
		\indent As a result of the above procedure, the following data was selected, sourced from national statistical offices and Eurostat:
		\begin{itemize}
			\item Consumer Price Index (CPI) – Monthly data from January 1991 to February 2025,
			\item Employment – Monthly data from January 2006 to January 2025,
			\item Gross Domestic Product (GDP), Real Terms – Quarterly data from January 2009 to October 2024,
			\item Gross Domestic Product (GDP), Nominal Terms – Quarterly data from January 2004 to October 2024,
			\item Population – Annual data from January 1949 to January 2020,
		\end{itemize}
		For data description with list of specific countries included see Appendix A.  Additionally, further data was selected from the Connect Platform, which is available through trusted data provider S\&P Global Market Intelligence:
		\begin{itemize}
			\item Wholesale Producer Price Index (WPPI), Nominal Terms – Annual data from January 1981 to January 2024, 
			\item Unemployment Rate, Quarterly Data – From April 1992 to October 2024, 
			\item Unemployment Rate, Annual Data – From January 1983 to January 2024, 
			\item Short-Term Interest Rates, Quarterly Data – From January 1996 to October 2024, 
			\item Short-Term Interest Rates, Quarterly Data – From January 1995 to October 2024, 
			\item Long-Term Interest Rates, Monthly Data – From May 1995 to December 2024, 
			\item Long-Term Interest Rates, Annual Data – From January 1971 to January 2024, 
		\end{itemize}
		For data description with list of specific countries included see Appendix A. In the CPI and Short Interest Rate datasets, we excluded Brazil due to a sharp increase in its values for some periods, which caused the forecast error to peak at approximately 50 times higher than the errors of the other countries. A similar situation occurred for Brazil and Peru in the Wholesale Producer Price Index, where the forecast error peak was around 10 times greater than that of other countries.\\
		\indent Additionally, a comprehensive set of Harmonized Index of Consumer Prices (HICP) indicators was sourced from Eurostat for EU countries. These indicators included numerous concepts, all rebased to 2015. For the list of concepts included see Appendix A. These indices were analysed over the period from December 2000 to December 2024, in accordance with the broader dataset constraints.\\
		\indent It is important to note that local economic policies and structural differences within countries may affect the quality and consistency of the data. This could impact the accuracy of forecasts, as models assume methodological consistency in the time series used for training and in future predictions.
		\indent For further analysis, we have also selected major stock market indices that serve as key indicators of financial market performance across different regions, sourced from Yahoo Finance. These include: 
		\begin{itemize}
			\item Dow Jones Industrial Average (DJIA) represents 30 large, publicly owned companies based in the United States and is often used as a barometer for the overall health of the U.S. economy,
			\item FTSE 100 refers to a broader index of major companies listed on the London Stock Exchange, reflecting the performance of the UK market,
			\item Hang Seng Index tracks the largest companies listed on the Hong Kong Stock Exchange and is an important measure of the Hong Kong and broader Chinese economy,
			\item Nasdaq Composite includes over 3,000 stocks listed on the Nasdaq Stock Market, with a strong emphasis on technology companies such as Apple, Microsoft, and Amazon,
			\item Nikkei 225 monitors 225 large, publicly owned companies on the Tokyo Stock Exchange and serves as a primary indicator of Japan’s stock market performance,
			\item S\&P 500 comprises 500 of the largest companies listed on U.S. stock exchanges and is one of the most accurate representations of the U.S. stock market.
		\end{itemize}
		We used daily data from 2021.07.14 until 2025.07.14.\\
		\indent We divided each dataset into a training and testing sets. Since one of the main features of DMS is it's ability to avoid the accumulation of errors in consecutive forecasts, a problem that often occurs in IMS, hence, the testing data set must be sufficiently large. The another reason for large testing set is that difference from IMS might be rather small in small forecasting horizons, as the first retraining is carried out after one year/month, and for that time differences in forecasts might still relatively small. After the second retraining, that is, in five years/months time, the difference should be more visible. On the other hand, there should be sufficient training data to properly train models. The balance was found to have around 10-15 years/months of trainig data and 20-25 years/months of testing data.
	\subsection{Comparison methodology}
	We measure the performance of the EDMS methodology by comparing it with the EIMS, as EIMS is a conventional forecasting methodology and serves as a natural benchmark. The comparison involves calculating the Mean Absolute Percentage Error (MAPE) for each time series in the datasets and take the average of that. This denoted as $\overline{mape}$. We selected MAPE because this metric is well-suited for comparing multiple time series and calculating an average metric. For instance, in the case of Mean Absolute Error (MAE), each time series may have values of different magnitudes, leading to considerable variation in MAE scores. Consequently, the contributions of some series to the overall average may distort the final value. In contrast, MAPE mitigates this issue since all errors are expressed on the same scale, that is, percentages.\\
	\indent We perform these computations separately for each dataset, and the expression for the average MAPE is as follows: 
		\begin{equation}
			\label{average_mape}
			\overline{mape}=\frac{1}{M}\sum_{j=1}^{M}\frac{1}{N}\sum_{h=1}^{H} \bigg|\frac{\widehat{y}_{t+h}^{j}-y_{t+h}^{j}}{\widehat{y}_{t+h}^{j}}\bigg|,
		\end{equation}
		where $M$ represents the number of time series for a particular dataset, $H$ is the number of observations in the forecasting horizon, $y_{t+h}^{j}$ are the testing observations of the $j$-th time series, and $t+h$ is the time step at which forecasts are computed. Here $\widehat{y}_{t+h}^{j}$ can be either, EDMS or EIMS forecast.\\
		\indent Recall that the forecast using EIMS methodology at time step $t+h$ can be expressed by the following formula: 
		\begin{equation*}
			\label{ensembling_eims}
			\widehat{y}_{t+h}=\sum_{i}w_{t,i}\widetilde{y}_{t+h,i}(Y_{t}),
		\end{equation*}
		where $w_{t,i}$ are the weights computed from the initial dataset, $h = 1,2,\ldots,H$ are the steps in the horizon, and $\widetilde{y}_{t+h,i}$ are the forecasts at a time step one year ahead from time $t$ for the $i$-th model, which is trained on the data up to time $t$, denoted as $Y_{t}=(y_{1}, y_{2}, \ldots, y_{N})$.\\
		\indent\qquad Furthermore, to clearly indicate the difference of EDMS over EIMS, we compute the percentage change in average MAPE for each dataset, defined as:
		\begin{equation*}
			\Delta(\%) = \frac{\overline{mape}_{ims} - \overline{mape}_{dms}}{\overline{mape}_{ims}}
		\end{equation*}
	\subsection{Results}
		In Tables \ref{tab:Comparison_MAPE_annual}, \ref{tab:Comparison_MAPE_quarter} and \ref{tab:Comparison_MAPE_month} 
we present the results of our computations for economic data. We categorize the results into annual, quarterly, and monthly frequencies due to the substantial differences in the amount of training data used for each frequency, which significantly impacts the analysis results.  In the tables, each row corresponds to a distinct dataset.  In each row, in addition to the average MAPE errors (\ref{average_mape}) for the EIMS and EDMS methodologies and the values of $\Delta(\%)$, we also list the number of time series, the total number of observations, and the number of observations for testing. At the bottom of each table, we have included the average percentage changes.\\
		\begin{table}[h]
			\begin{center}
				\begin{tabular}{ | p{5,7cm}| l | l | l | l | l |}
					\hline
					Data set & Number of & Total/test & EIMS & EDMS & $\Delta(\%)$\\ 
					& time series & data size & & & \\ 
					\hline
					Long Interest Rates & 21 & 54/35 & 6,51 & 3,21 & \textbf{50,73} \\ 
					Short Interest Rates & 19 & 51/35 & 9,61 & 5,79 & \textbf{39,73} \\ 
					Production Industries & 41 & 45/30 & 0,19 & 0,17 & \textbf{9,76} \\
					Unemployment & 32 & 42/30 & 1,06 & 0,66 & \textbf{37,91}  \\ 
					Employment & 16 & 53/35 & 0,08 & 0,07 & \textbf{-1,98} \\ 
					Population & 12 & 72/50 & 0,08 & 0,06 & \textbf{16,00} \\ 
					Wholesale Producer Price Index  & 38 & 44/30 & 15,59 & 8,46 & \textbf{45,70} \\ 
					Average $\Delta(\%)$ & & & & & \textbf{28,26}  \\ \hline
				 \end{tabular}
			\end{center}
			\caption{Comparison of MAPE errors for EDMS and EIMS, for data sets with annual frequency}
			\label{tab:Comparison_MAPE_annual}
		\end{table}
		\begin{table}[h]
			\begin{center}
				\begin{tabular}{ | p{4,7cm}| l | l | l | l | l |}
					\hline
					Data set & Number of & Total/test & EIMS & EDMS & $\Delta(\%)$\\ 
					& time series & data size & & & \\ 
					\hline
					GDP Nominal & 10 & 84/55 & 0,89 & 0,32 & \textbf{64,21}  \\ 
					GDP Real & 8 & 64/40 & 0,038 & 0,033 & \textbf{12,24}  \\ 
					Long Interest Rates & 40 & 120/85 & 8,90 & 3,93 & \textbf{55,80} \\ 
					Short Interest Rates & 82 & 116/80 & 5,89 & 3,02 & \textbf{48,81} \\ 
					Production Industries & 64 & 124/85 & 0,16 & 0,14 & \textbf{10,76} \\ 
					Unemployment & 49 & 131/90 & 1,19 & 0,84 & \textbf{29,20} \\ 
					Average $\Delta(\%)$ & & & & & \textbf{36,84}  \\ \hline
				 \end{tabular}
			\end{center}
			\caption{Comparison of MAPE errors for EDMS and EIMS, for data sets with quarterly frequency}
			\label{tab:Comparison_MAPE_quarter}
		\end{table}
		\begin{table}[h]
			\begin{center}
				\begin{tabular}{ | p{4,7cm}| l | l | l | l | l |}
					\hline
					Data set & Number of & Total/test & EIMS & EDMS & $\Delta(\%)$\\ 
					& time series & data size & & & \\ 
					\hline
					Employment & 5 & 229/170 & 10,65 & 3,46 & \textbf{67,52}  \\ 
					CPI & 13 & 410/300 & 2,30 & 0,85 & \textbf{63,06}  \\ 
					CPI by categories & 497 & 289/210 & 0,07  & 0,06 & \textbf{14,41}  \\
					Long Interest Rates & 30 & 356/280 & 8,05 & 3,58 & \textbf{55,52} \\ 
					Short Interest Rates & 79 & 348/270 & 9,91 & 3,99 & \textbf{59,76} \\ 
					Production Industries & 58 & 324/255 & 0,23 & 0,16 & \textbf{30,89} \\ 
					Average $\Delta(\%)$ & & & & & \textbf{48,53}  \\ \hline
				 \end{tabular}
			\end{center}
			\caption{Comparison of MAPE errors for EDMS and EIMS, for data sets with monthly frequency}
			\label{tab:Comparison_MAPE_month}
		\end{table}
	\indent The average improvement differ significantly across frequencies, averaging 28\% for annual frequency, 36\% for quarterly, and 48\% for monthly. These discrepancies, although substantial, follow a pattern: the accuracy increases linearly by approximately 10\% with each increase in frequency. Notably, the increase in frequency corresponds to a rise in the amount of data provided. For instance, the annual frequency typically has around 50 data points, quarterly frequency has about 100 observations, and monthly frequency has approximately 300. Similarly, the testing data consists of around 30, 80, and 240 observations for annual, quarterly, and monthly frequencies, respectively. This translates to an approximate forecasting horizon of 30 years for annual data, decreasing to 20 years for both quarterly and monthly data. we can notice that the increase in the amount of data is significantly steeper than the improvement in accuracy. The impact of this steep increase on performance may be mitigated by the fact that actual forecasting periods are becoming shorter, as EDMS forecasts tend to perform better for longer horizons than for shorter ones. Therefore, it is reasonable to conclude that we observe a linear increase in overall performance.\\
	\indent The results show that the EDMS significantly outperforms the EIMS for almost every data set. This is particularly evident in high-variability datasets such as CPI, where EDMS achieved significantly better results. CPI is influenced by monthly price changes, often driven by energy, food, and housing costs. These can be subject to shocks (e.g., oil price changes). Similarly, for GDP (real and nominal), EDMS provides more accurate forecasts. The only exception was Employment (annual), where EIMS slightly outperformed EDMS, probably due to the smoother and more predictable nature of demographic trends which benefit from iterative method. Economic time series like interest rates and production often exhibit non-linear trends. For example, changes in interest rates may affect investment and employment with a delay, and these effects may not be proportional. EDMS is better in handling these behaviours because it integrates the smoothing process over time, allowing it to adapt to long-term trends and gradual changes more effectively, e.g. interest rates often follow long cycles influenced by monetary policy and global capital flows. Production and prices can be affected by supply chain innovations or external shocks (e.g. commodity shocks/unexpected tariffs). In contrast, unemployment rate is influenced by structural factors like labour market reforms or demographic changes. EDMS can accommodate these slow-moving shifts, while EIMS might miss them or misinterpret them as noise. Employment data at monthly frequency reflects both seasonal hiring patterns and long-term labour market trends. These patterns can be irregular due to holidays, school cycles, or policy changes, so here also EDMS can be advantageous.\\
	\indent The potential impact of the number of time series on the comparison is relatively negligible, as, aside from a few datasets with slightly fewer time series than average, specifically GDP, Monthly Employment, Population, and CPI, the number of time series tends to be roughly the same within each frequency group.\\
	\begin{table}[h]
		\begin{center}
			\begin{tabular}{ | p{5,4cm}| l | l | l | l | l |}
				\hline
				Index & EIMS & EDMS & $\Delta(\%)$\\ 
				\hline
				Dow Jones Industrial Average & 0,26 & 0,2 & \textbf{23,57} \\ 
				FTSE 500 & 0,14 & 0,11 & \textbf{22,94} \\ 
				Hang Seng & 0,26 & 0,17 & \textbf{32,89} \\ 
				Nasdaq Composite & 0,39 & 0,28 & \textbf{25,89}  \\ 
				Nikkei 225 & 0,14 & 0,15 & \textbf{-6,7}  \\ 
				S\&P 500 & 0,31 & 0,24 & \textbf{24,37} \\ 
				Average $\Delta(\%)$ & & & \textbf{20,49}  \\ \hline
			 \end{tabular}
		\end{center}
		\caption{Comparison of MAPE errors for EDMS and EIMS, for 6 indices with daily frequency. Entire dataset has 1041 observations and 720 testing observations}
		\label{tab:Comparison_MAPE_daily}
	\end{table}
	\indent In Table \ref{tab:Comparison_MAPE_daily}, we present the results of computations for financial data, where each row represents a distinct index, although the models were trained on data from all indices. Each table's row shows the average MAPE errors (\ref{average_mape}) for the EIMS and EDMS methodologies, as well as the values of $\Delta(\%)$. The total number of observations and the number of observations for testing are consistent across all indices, with 1041 observations in total, 720 observations used for testing, and remaining for training. Additionally, we have included the average of percentage changes at the bottom of the table.\\
	\indent We observe a significant improvement of accuracy for financial data, averaging at 20\%. However, in these experiments, we do not see the continuation of the pattern where an increase in the amount of data leads to improved performance. This time, the improvement is similar to that observed with annual economic data. This irregularity can be explained by the fact that we are only considering six time series of a similar category. Consequently, the sample for the average is small and homogeneous. Moreover, each index is similarly affected by major crises and other global events. Such a sample may not accurately represent the true performance of the methodology. In contrast, in the experiments with economic data, each average was derived from around six heterogeneous data sets, where even the time series within each data set could differ substantially, as they represented different countries, often from various continents.\\
	\indent Another reason for the disrupted pattern of performance improvement with the amount of data, is the substantially different nature of financial data. Specifically, the data is daily, and the market is more liquid. Furthermore, there is considerable noise in the data and potential for speculation, and indices react strongly and promptly to various types of disruptions, such as tariffs, i.e., shocks.\\
	\indent Additionally, we compute the average percentage improvement across the entire sample of data sets, which is equal to 33.32\%. This significant improvement in overall accuracy supports the assertion made in the description of EDMS. Specifically, that the methodology not only leverages the strengths of Ensemble and DMS methods but is also highly influenced by their synergistic effect. The synergy can be inferred from the fact that applying DMS alone to EIMS would not be as beneficial. This is because DMS generally provides only slight improvements, and not in all cases, as demonstrated in previous works on DMS discussed in the literature review.

\section{Conclusions}
\subsection{Results of comparative analysis}
In light of the extensive comparative analysis conducted for the newly introduced Ensembled Direct Multi-Step (EDMS) methodology, we can confirm our theoretical claim. The average improvement across all datasets was 33.32\% in favour of EDMS. This suggests that incorporating just a few retrainings into the iterative ensembling process yields significantly better accuracy, regardless of the frequency of considered economic indicators. Indeed, the only plausible explanation for the observed improvement in accuracy is the synergy between the Ensembling technique and DMS, resulting in an overall enhancement that surpasses the benefits of each method when applied individually.\\

\indent The improvement for macroeconomic indicators follows a clear trend based on data frequency, with annual data averaging around 28\%, quarterly data at approximately 36\%, and monthly data reaching about 48\%. Notably, the increase in the amount of data with increase of frequencies is much steeper than the corresponding improvement in accuracy. The steepness of the data size impact may be mitigated by shorter forecasting periods for higher frequencies, as EDMS forecasts should demonstrate improved performance over longer horizons. Thus, a linear increase in overall performance is reasonable.\\
\indent From an economic perspective, considered time series represent key macroeconomic indicators that help policymakers assess and forecast the direction of the economy. Economic relationships are rarely linear or static. They evolve over time, influenced by policy decisions, expectations, and external shocks. EDMS’s architecture is designed to adapt and retain long-term dependencies, which mirrors how past crises, policy changes, or demand shocks continue to influence current economic behaviour. This memory aspect is important factor in economics, where historical context often shapes future expectations and actions. EDMS models are inherently designed at capturing these nonlinear and dynamic patterns.

\indent A significant improvement of 20\% is observed for financial data. However, this performance increase does not align with the pattern of improvement seen with increasing amount of data for macroeconomic data. This irregularity may arise from the use of a significantly smaller and more homogeneous sample of time series. Moreover, financial data is substantially different from economic data, notably it is more variable and less predictable in nature. Therefore, since model specification is more demanding, theoretically, EDMS should perform even better than EIMS. This fact strengthens the claim that size of time series sample might have been insufficient.

\indent Results of comparative analysis indicate that EDMS can be successfully applied to forecasting major stock market indices. Financial indices like the Dow Jones, Nasdaq, FTSE, Hang Seng, Nikkei, and S\&P 500 exhibit high-frequency volatility, driven by market sentiment, global events and investor behavior. These series are typically noisy, with strong fluctuations and less predictable long-term trends compared to macroeconomic indicators. EDMS adapts better to short-term patterns in volatile data. However, Nikkei 225 EDMS performs slightly worse than EIMS. This suggests that for certain indices, especially those with less noise or more stable behaviour, the simpler method may be sufficient or even preferable.

\indent Intuitively, our comparative analysis indicates that adding more data during the training process is more beneficial than not adding data at all. This, combined with dynamically changing weights in the ensemble with each data addition, results in significant improvements in accuracy.

\subsection{Improvements and Extensions}
There are several ways to improve or extend the proposed methodology. A simple and natural enhancement is to increase the number of retrainings. Essentially, this introduces more data into the process and should further increase accuracy. This is aligned with common data augmentation practices in machine learning, such as generating additional synthetic data to improve training. The main cost of such an improvement would be an increase in computational load, which might acceptable for some use cases.

\indent The improvement in forecast accuracy is particularly noticeable for medium- and long-term horizons. In our experiments, this effect became evident after a few years for economic data and after a few months for financial data. Given sufficient data frequency and volume, it is possible to shift the retraining dates to earlier horizons, allowing the accuracy gains to manifest sooner. For instance, if the improvement becomes visible after a few months using monthly data, with retrainings at third and fifteenth months.  Given the daily data is available, we could set retraining intervals at first and fifth weeks for daily data. This adjustment may enable the improvement effect to emerge within a few weeks instead of months.

\indent The literature offers various enhancements to the DMS procedure and Ensemble techniques, some of which were discussed in the Literature Review section. In theory, any such improvement should be equally applicable to the proposed methodology.

In the future, it may be beneficial to incorporate additional statistical and machine learning models into the EDMS framework, potentially enhancing its robustness and predictive power.

\subsection{Potential applications}
The proposed approach can be naturally applied to forecasting other macroeconomic variables as well as other types of financial data. Importantly, there is nothing in the methodology that restricts its use to macroeconomic or financial data. Therefore, it is worth exploring the applicability of the proposed approach in other domains, such as healthcare (e.g. disease outbreak prediction), environmental science (e.g. weather and climate modelling), transportation (e.g. traffic flow and delivery time estimation), energy (e.g. load and generation prediction, renewable energy generation prediction).
\indent Based on our research, this methodology is particularly well-suited for improving medium- and long-term economic and financial forecasting. As such, it holds strong potential for use in institutions where accurate mid-long term forecasting is critical, including: central banks or other regulatory institutions, to support monetary policy decisions and promote economic stability, governments to inform fiscal policy, assess macroeconomic trends, and develop strategies for sustainable growth. Additionally, commercial banks and other financial institutions can leverage EDMS to enhance risk assessment, improve asset management strategies, and optimize investment decisions. Investment firms and individual investors may also find value in using the methodology to predict market trends and identify emerging opportunities, thereby enabling more informed decision-making. \\

\begin{contributions}
authors have the following contributions:\\
Tomasz M. \L api\'nski - designed and described the methodology, developed related software and conducted experiments, designed and described comparative analysis along with results, reviewed in detail and consolidated content of the manuscript,\\
Krzysztof Zi\'o\l kowski - wrote introduction and conclusions, explained economic context in results, conducted and wrote literature review, gathered, prepared and described data for experiments, reviewed the manuscript,\\
\end{contributions}

\begin{acknowledgments}
	We are incredibly grateful to David Douglas Willson from S\&P Global Market Intelligence for his consistent support in this work. This includes creating the preconditions for this research, making available the algorithms for ensemble forecasting, including Average Growth, Linear and Polynomials models, and for help in developing the software for ETS and LSTM models. Additionally, for his comment about the insensitivity of regression models to retraining.
\end{acknowledgments}

\begin{appendices}
\section{Appendix}
	Detailed description of data, including countries sourced from national statistical offices and Eurostat:
	\begin{itemize}
		\item Consumer Price Index (CPI) – Monthly data from January 1991 to February 2025, covering: Argentina, Australia, India, Mainland China, Germany, Brazil, South Africa, Poland, France, Japan, and the United Kingdom.
		\item Employment – Monthly data from January 2006 to January 2025, covering: Australia, Germany, Japan, Brazil, and Poland.
		\item Gross Domestic Product (GDP), Real Terms – Quarterly data from January 2009 to October 2024, covering: Germany, Poland, France, Japan, Brazil, Argentina, United Kingdom, and South Africa.
		\item Gross Domestic Product (GDP), Nominal Terms – Quarterly data from January 2004 to October 2024, covering: South Africa, United Kingdom, Mainland China, Argentina, Brazil, Poland, Japan, Germany, Australia, and France.
		\item Population – Annual data from January 1949 to January 2020, covering: Brazil, Poland, Mainland China, India, Japan, Argentina, United Kingdom, Germany, Australia, South Africa, and France.\\
	\end{itemize}
	Detailed description of data, including countries, sourced from the Connect Platform, which is available through S\&P Global Market Intelligence:
	\begin{itemize}
		\item Wholesale Producer Price Index (WPPI), Nominal Terms – Annual data from January 1981 to January 2024, covering: Australia, Austria, Belgium, Bhutan, Canada, Chile, Colombia, Denmark, Egypt, Germany, Greece, Hungary, Indonesia, Ireland, Israel, Japan, Jordan, Liechtenstein, Luxembourg, Mexico, Morocco, Netherlands, New Zealand, Nigeria, Norway, Singapore, South Africa, South Korea, Spain, Sweden, Switzerland, Taiwan, Turkey, United Kingdom, United States, and Uruguay.
		\item Unemployment Rate, Quarterly Data – From April 1992 to October 2024, covering: Angola, Argentina, Australia, Belgium, Brazil, Bulgaria, Canada, Chile, Colombia, Denmark, Egypt, Finland, France, Germany, Greece, Hong Kong SAR, Hungary, India, Indonesia, Ireland, Israel, Italy, Japan, Kenya, Luxembourg, Malaysia, Mexico, Netherlands, New Zealand, Norway, Philippines, Poland, Portugal, Qatar, Russia, Singapore, South Africa, South Korea, South-East Asia, Spain, Sri Lanka, Sweden, Switzerland, Taiwan, Thailand, Turkey, United Kingdom, United States, and Uruguay.
		\item Unemployment Rate, Annual Data – From January 1983 to January 2024, covering: Australia, Belgium, Brazil, Canada, Colombia, France, Germany, Greece, Hong Kong SAR, India, Ireland, Israel, Italy, Japan, Luxembourg, Malaysia, Netherlands, Norway, Panama, Philippines, Portugal, Puerto Rico, South Africa, South Korea, Spain, Sweden, Switzerland, Taiwan, Thailand, Turkey, United Kingdom, and United States.
		\item Short-Term Interest Rates, Quarterly Data – From January 1996 to October 2024, covering: Algeria, Americas, Angola, Argentina, Australia, Austria, Bahrain, Bangladesh, Belgium, Botswana, Brazil, Bulgaria, Canada, Chile, China (mainland), Colombia, Costa Rica, Croatia, Cyprus, Czechia, Denmark, Dominican Republic, Ecuador, Egypt, Estonia, Finland, France, Germany, Ghana, Greece, Guatemala, Hong Kong SAR, Hungary, Iceland, India, Indonesia, Ireland, Israel, Italy, Japan, Kazakhstan, Kenya, Kuwait, Lebanon, Levant, Lithuania, Luxembourg, Malta, Mexico, Morocco, Namibia, Netherlands, New Zealand, Nigeria, North Africa, Norway, Oman, Paraguay, Peru, Philippines, Poland, Portugal, Qatar, Romania, Russia, Saudi Arabia, Slovakia, South Africa, South Korea, Spain, Sri Lanka, Sweden, Switzerland, Taiwan, Thailand, Tunisia, Turkey, Uganda, United States, Uruguay, Venezuela, and Zambia.
		\item Short-Term Interest Rates, Quarterly Data – From January 1995 to October 2024, covering: Argentina, Australia, Austria, Belgium, Bulgaria, Canada, Croatia, Denmark, Estonia, Finland, France, Germany, Greece, Hungary, Iceland, Ireland, Israel, Italy, Japan, Luxembourg, Malaysia, Mexico, Namibia, Netherlands, New Zealand, Nigeria, Norway, Philippines, Portugal, Slovakia, South Africa, South Korea, Spain, Sweden, Switzerland, Taiwan, Thailand, Turkey, United Kingdom, and United States.
		\item Long-Term Interest Rates, Monthly Data – From May 1995 to December 2024, covering: Argentina, Australia, Austria, Belgium, Denmark, Estonia, Eurozone, Finland, France, Germany, Iceland, Ireland, Italy, Japan, Luxembourg, Malaysia, Namibia, Netherlands, New Zealand, Nigeria, Norway, South Africa, South Korea, Spain, Sweden, Switzerland, Taiwan, Thailand, United Kingdom, and United States.
		\item Long-Term Interest Rates, Annual Data – From January 1971 to January 2024, covering: Andorra, Argentina, Australia, Austria, Belgium, Canada, Denmark, France, Germany, Ireland, Italy, Japan, Netherlands, New Zealand, Nigeria, Portugal, South Africa, Sweden, Switzerland, United Kingdom, and United States.
	\end{itemize}
	List of concepts for Harmonized Index of Consumer Prices (HICP) indicators sourced from Eurostat for EU countries:
	\begin{itemize}
		\item All-Items excluding Administered Prices
		\item Overall Index excluding Tobacco
		\item All-Items excluding Mainly Administered Prices
		\item Overall Index excluding Energy and Seasonal Food
		\item Overall Index excluding Alcohol and Tobacco
		\item Overall Index excluding Frequent Out-of-Pocket Purchases
		\item Overall Index excluding Education, Health, and Social Protection
		\item Overall Index excluding Energy and Unprocessed Food
		\item Overall Index excluding Energy, Food, Alcohol, and Tobacco
		\item Overall Index excluding Energy
		\item Overall Index excluding Seasonal Food
		\item Overall Index excluding Housing, Water, Electricity, Gas, and Other Fuels
		\item Total Index
		\item Overall Index excluding Liquid Fuels and Lubricants for Personal Transport Equipment
		\item All-Items excluding Fully Administered Prices
	\end{itemize}

\end{appendices}


\bibliographystyle{apa}
\bibliography{biblio}

\begin{thebibliography}{}

\bibitem[\protect\astroncite{pro}{2024}]{prophet_github}
 (2024).
\newblock Prophet | forecasting at scale.
\newblock GitHub - facebook/prophet: Tool for producing high quality forecasts
  for time series data that has multiple seasonality with linear or non-linear
  growth.

\bibitem[\protect\astroncite{Ahmed
  et~al.}{2010}]{ahmed_empirical_comparison_ml_models}
Ahmed, N., Atiya, A., Gayar, N., and El-Shishiny, H. (2010).
\newblock An empirical comparison of machine learning models for time series
  forecasting.
\newblock {\em Economic Review}, 29(5–6):594–621.

\bibitem[\protect\astroncite{Albahli}{2025}]{albahli_lstm_vs_prophet}
Albahli, S. (2025).
\newblock Lstm vs. prophet: Achieving superior accuracy in dynamic electricity
  demand forecasting.
\newblock {\em Energies}, 18(2):278.

\bibitem[\protect\astroncite{Aziz et~al.}{2022}]{aziz_facebook_prophet}
Aziz, M., Barawi, M., and Shahiri, H. (2022).
\newblock Is facebook prophet superior than hybrid arima model to forecast
  crude oil price?
\newblock Technical report, Universiti Kebangsaan Malaysia.

\bibitem[\protect\astroncite{Azqueta-Gavaldón}{2017}]{azqueta_news_2017}
Azqueta-Gavaldón, A. (2017).
\newblock Developing news-based economic policy uncertainty index with
  unsupervised machine learning.
\newblock {\em Economics Letters}, 158:47--50.

\bibitem[\protect\astroncite{Bates and Granger}{1969}]{bates_combination_1969}
Bates, J.~M. and Granger, C. W.~J. (1969).
\newblock The combination of forecasts.
\newblock {\em Operational Research Quarterly}, 20(4):451--468.

\bibitem[\protect\astroncite{Bhansali}{1997}]{bhansali_direct_autoregressive_predictors}
Bhansali, R.~J. (1997).
\newblock Direct autoregressive predictors for multistep prediction: Order
  selection and performance relative to plug-in predictors.
\newblock {\em Statistica Sinica}, 7(2):425--449.

\bibitem[\protect\astroncite{Booth et~al.}{2014}]{booth_automated_2014}
Booth, A., Gerding, E., and McGroarty, F. (2014).
\newblock Automated trading with performance weighted random forests and
  seasonality.
\newblock {\em Expert Systems with Applications}, 41(8):3651--3661.

\bibitem[\protect\astroncite{Breiman}{1996}]{ensemble_3}
Breiman, L. (1996).
\newblock {Stacked regressions}.
\newblock {\em Machine Learning}, 24(1):49--64.

\bibitem[\protect\astroncite{Brykin}{2024}]{brykin_sales_forecasting_models}
Brykin, D. (2024).
\newblock Sales forecasting models: Comparison between arima, lstm and prophet.
\newblock {\em Journal of Computer Science}, 20(10):1222–1230.

\bibitem[\protect\astroncite{Chen
  et~al.}{2024}]{chen_time_series_forecasting_oil_production}
Chen, G., Tian, H., Xiao, T., Xu, T., and Lei, H. (2024).
\newblock Time series forecasting of oil production in enhanced oil recovery
  system based on a novel cnn-gru neural network.
\newblock {\em Geoenergy Science and Engineering}, 233.

\bibitem[\protect\astroncite{Chen et~al.}{2007}]{chen_flexible_2007}
Chen, Y., Yang, B., and Abraham, A. (2007).
\newblock Flexible neural trees ensemble for stock index modeling.
\newblock {\em Neurocomputing}, 70(4):697--703.

\bibitem[\protect\astroncite{Cheng
  et~al.}{2024}]{cheng_forecasting_bitcoin_prices}
Cheng, J., Tiwari, S., Khaled, D., Mahendru, M., and Shahzad, U. (2024).
\newblock Forecasting bitcoin prices using artificial intelligence: combination
  of ml, sarima, and facebook prophet models.
\newblock {\em Technological Forecasting and Social Change}, 198.

\bibitem[\protect\astroncite{Chevillon}{2007}]{chevillon_direct_multi_step}
Chevillon, G. (2007).
\newblock Direct multi-step estimation and forecasting.
\newblock {\em Journal of Economic Surveys}, 21(4):746--785.

\bibitem[\protect\astroncite{Chevillon and
  Hendry}{2004}]{chevillon_non_parametric_direct_multi_step}
Chevillon, G. and Hendry, D. (2004).
\newblock Non-parametric direct multi-step estimation for forecasting economic
  processes.
\newblock Technical report, Department of Economics.
\newblock Discussion Paper Series, Oxford OX1 3UQ.

\bibitem[\protect\astroncite{Chris~Chatfield}{2019}]{time_series_analysis_1}
Chris~Chatfield, H.~X. (2019).
\newblock {\em The Analysis of Time Series: An Introduction with R, Seventh
  Edition}.
\newblock CRC Press.

\bibitem[\protect\astroncite{Clark and
  West}{2007}]{clark_west_approx_normal_tests}
Clark, T. and West, K. (2007).
\newblock Approximately normal tests for equal predictive accuracy in nested
  models.
\newblock {\em Journal of Econometrics}, 138:291--311.

\bibitem[\protect\astroncite{Cox}{1961}]{cox_prediction_exponentially_weighted}
Cox, D. (1961).
\newblock Prediction by exponentially weighted moving averages and related
  methods.
\newblock {\em Journal of the Royal Statistical Society}, B23:414--422.

\bibitem[\protect\astroncite{Dossani}{2022}]{dossani_inference_multi_step_forecasting}
Dossani (2022).
\newblock Inference in direct multi-step and long horizon forecasting
  regressions.
\newblock Available at SSRN: https://ssrn.com/abstract=4020488 or
  http://dx.doi.org/10.2139/ssrn.4020488.

\bibitem[\protect\astroncite{Findley}{1983}]{findley_use_multiple_models}
Findley, D. (1983).
\newblock On the use of multiple models for multi-period forecasting.
\newblock In {\em Proceedings of Business and Economic Statistics}, page
  528–531. American Statistical Association.

\bibitem[\protect\astroncite{Ghosh and
  Jana}{2024}]{ghosh_clean_energy_stock_price_forecasting}
Ghosh, I. and Jana, R. (2024).
\newblock Clean energy stock price forecasting and response to macroeconomic
  variables: A novel framework using facebook's prophet, neuralprophet and
  explainable ai.
\newblock {\em Technological Forecasting and Social Change}, 200.

\bibitem[\protect\astroncite{Gooijer and
  Hyndman}{2006}]{de_gooijer_25_years_forecasting}
Gooijer, J.~D. and Hyndman, R. (2006).
\newblock 25 years of time series forecasting.
\newblock {\em International Journal of Forecasting}, 22(3):443–473.

\bibitem[\protect\astroncite{Granger~C.W.J.}{1986}]{time_series_analysis_2}
Granger~C.W.J., N.~P. (1986).
\newblock {\em Forecasting Economic Time Series, Second Edition}.
\newblock Elsevier.

\bibitem[\protect\astroncite{Gumus and
  Kiran}{2017}]{gumus_crude_oil_forecasting}
Gumus, M. and Kiran, M. (2017).
\newblock Crude oil price forecasting using xgboost.
\newblock In {\em International Conference on Computer Science and Engineering
  (UBMK)}.

\bibitem[\protect\astroncite{Guruge and
  Priyadarshana}{2025}]{guruge_time_series_forecasting_kubernetes}
Guruge, P. and Priyadarshana, Y. (2025).
\newblock Time series forecasting-based kubernetes autoscaling using facebook
  prophet and long short-term memory.
\newblock {\em Frontiers in Computer Science}, 7:1509165.

\bibitem[\protect\astroncite{Göçken et~al.}{2016}]{gocken_integrating_2016}
Göçken, M., Özçalıcı, M., Boru, A., and Dosdoğru, A.~T. (2016).
\newblock Integrating metaheuristics and artificial neural networks for
  improved stock price prediction.
\newblock {\em Expert Systems with Applications}, 44:320--331.

\bibitem[\protect\astroncite{Hamarashid}{2021}]{lstm_images}
Hamarashid, H.~K. (2021).
\newblock {Modified Long Short-Term memory and Utilizing in Building sequential
  model}.
\newblock {\em International Journal of Multidisciplinary and Current
  Research}, 9(May-June):207--211.

\bibitem[\protect\astroncite{In and Jung}{2022}]{in_jung_simple_averaging}
In, J.-Y. and Jung, Y. (2022).
\newblock Simple averaging of direct and recursive forecasts via partial
  pooling using machine learning.
\newblock {\em International Journal of Forecasting}, 38(4):1386--1399.

\bibitem[\protect\astroncite{Jiang
  et~al.}{2025}]{jiang_trend_pattern_fuzzy_lstm}
Jiang, Y., Yu, F., Tang, Y., Ouyang, C., and Li, F. (2025).
\newblock Trend-pattern unlimited fuzzy information granule-based lstm model
  for long-term time-series forecasting.
\newblock {\em International Journal of Approximate Reasoning}, 180.

\bibitem[\protect\astroncite{Johnston
  et~al.}{1974}]{johnston_estimation_prediction_dynamic_models}
Johnston, H., Klein, L., and Shinjo, K. (1974).
\newblock Estimation and prediction in dynamic econometric models.
\newblock In Sellekaerts, W., editor, {\em Econometrics and Economic Theory}.
  Palgrave Macmillan.

\bibitem[\protect\astroncite{Joosery and Deepa}{2019}]{lstm_arima_comparison_3}
Joosery, B. and Deepa, G. (2019).
\newblock {Comparative analysis of time-series forecasting algorithms for stock
  price prediction}.
\newblock {\em Conference: AISS 2019: 2019 International Conference on Advanced
  Information Science and System}, (33):1--6.

\bibitem[\protect\astroncite{Kang}{2003}]{kang_multiperiod_forecasting}
Kang, I.-B. (2003).
\newblock Multiperiod forecasting using different models for different
  horizons: an application to u.s. economic time-series data.
\newblock {\em International Journal of Forecasting}, 19:387–400.

\bibitem[\protect\astroncite{Kashpruk}{2020}]{kashpruk_comparative_research_statistical_models}
Kashpruk, N. (2020).
\newblock {\em Comparative Research of Statistical Models and Soft Computing
  for Identification of Time Series and Forecasting}.
\newblock PhD thesis.

\bibitem[\protect\astroncite{Klein}{1971}]{klein_essay_theory_economics_prediction}
Klein, L. (1971).
\newblock An essay on the theory of economics prediction.
\newblock {\em The Journal of Finance}, 26(4):1010--1012.

\bibitem[\protect\astroncite{Kourentzes
  et~al.}{2014a}]{kourentzes_neural_2014b}
Kourentzes, N., Barrow, D.~K., and Crone, S.~F. (2014a).
\newblock Neural network ensemble operators for time series forecasting.
\newblock {\em Expert Systems with Applications}, 41:4235--4244.

\bibitem[\protect\astroncite{Kourentzes
  et~al.}{2014b}]{kourentzes_improving_2014a}
Kourentzes, N., Petropoulos, F., and Trapero, J.~R. (2014b).
\newblock Improving forecasting by estimating time series structural components
  across multiple frequencies.
\newblock {\em International Journal of Forecasting}, 30(2):291--302.

\bibitem[\protect\astroncite{Kristjanpoller
  et~al.}{2014}]{kristjanpoller_volatility_2014}
Kristjanpoller, W., Fadic, A., and Minutolo, M.~C. (2014).
\newblock Volatility forecast using hybrid neural network models.
\newblock {\em Expert Systems with Applications}, 41(5):2437--2442.

\bibitem[\protect\astroncite{Kumar and
  Lyngdoh}{2020}]{kumar_prediction_area_production}
Kumar, S. and Lyngdoh, F. (2020).
\newblock Prediction of area and production of groundnut using box-jenkins
  arima and neural network approach.
\newblock {\em Journal of Reliability and Statistical Studies}.

\bibitem[\protect\astroncite{Kwarteng and
  Andreevich}{2024}]{kwarteng_comparative_analysis_arima_sarima_prophet}
Kwarteng, S. and Andreevich, P. (2024).
\newblock Comparative analysis of arima, sarima and prophet model in
  forecasting antidiabetic drug demand in australia.
\newblock {\em Research \& Development}, 5(4):123–130.

\bibitem[\protect\astroncite{Li
  et~al.}{2005}]{li_recent_developments_econometric_modeling}
Li, G., Song, H., and Witt, S. (2005).
\newblock Recent developments in econometric modeling and forecasting.
\newblock {\em Journal of Travel Research}, 44(1):82–99.

\bibitem[\protect\astroncite{Li
  et~al.}{2025}]{li_enhancing_financial_forecasting}
Li, J.-C., Sun, L.-P., Wu, X., and Tao, C. (2025).
\newblock Enhancing financial time series forecasting with hybrid deep
  learning: Ceemdan-informer-lstm model.
\newblock {\em Applied Soft Computing}, 177.

\bibitem[\protect\astroncite{Li
  et~al.}{2022}]{li_time_series_production_forecasting}
Li, X., Ma, X., Xiao, F., Xiao, C., Wang, F., and Zhang, S. (2022).
\newblock Time-series production forecasting method based on the integration of
  bidirectional gated recurrent unit (bi-gru) network and sparrow search
  algorithm (ssa).
\newblock {\em Journal of Petroleum Science and Engineering}, 208(Part A).

\bibitem[\protect\astroncite{Lin and
  Tsay}{1996}]{lin_tsay_co_integration_constraint}
Lin, J. and Tsay, R. (1996).
\newblock Co-integration constraint and forecasting: an empirical examination.
\newblock {\em Journal of Applied Econometrics}, 11:519–538.

\bibitem[\protect\astroncite{M.~Pirani et~al.}{2022}]{lstm_arima_comparison_2}
M.~Pirani, P.~Thakkar, P.~J., Bohara, M.~H., and Garg, D. (2022).
\newblock {A Comparative Analysis of ARIMA, GRU, LSTM and BiLSTM on Financial
  Time Series Forecasting}.
\newblock {\em 2022 IEEE International Conference on Distributed Computing and
  Electrical Circuits and Electronics}, pages 1--6.

\bibitem[\protect\astroncite{Marcellino
  et~al.}{2006a}]{marcellino_comparison_direct_iterated}
Marcellino, M., Stock, J., and Watson, M. (2006a).
\newblock A comparison of direct and iterated multistep ar methods for
  forecasting macroeconomic time series.
\newblock {\em Journal of Econometrics}, 135(1–2):499--526.

\bibitem[\protect\astroncite{Marcellino
  et~al.}{2006b}]{marcellino_comparison_direct_iterated_micro}
Marcellino, M., Stock, J., and Watson, M. (2006b).
\newblock A comparison of direct and iterated multistep ar methods for
  forecasting microeconomic time series.
\newblock {\em Journal of Econometrics}, 135:499–526.

\bibitem[\protect\astroncite{McCracken and
  McGillicuddy}{2019}]{mccracken_empirical_investigation_forecasts}
McCracken, M.~W. and McGillicuddy, J.~T. (2019).
\newblock An empirical investigation of direct and iterated multistep
  conditional forecasts.
\newblock {\em Journal of Applied Econometrics}, 34(2):181–204.

\bibitem[\protect\astroncite{McElroy}{2015}]{mcelroy_direct_iterative_identical}
McElroy, T. (2015).
\newblock When are direct multi-step and iterative forecasts identical.
\newblock {\em Journal of Econometrics}, 34(4):315–336.

\bibitem[\protect\astroncite{Newbold and
  Granger}{1974}]{newbold_experience_1974}
Newbold, P. and Granger, C. W.~J. (1974).
\newblock Experience with forecasting univariate time series and the
  combination of forecasts.
\newblock {\em Journal of the Royal Statistical Society. Series A (General)},
  137(2):131--165.

\bibitem[\protect\astroncite{O.B.~Sezer}{2020}]{systematic_review}
O.B.~Sezer, M.U.~Gudelek, A.~O. (2020).
\newblock Financial time series forecasting with deep learning: A systematic
  literature review: 2005–2019.
\newblock {\em Applied Soft Computing Journal}, 90(May):106181.

\bibitem[\protect\astroncite{Opitz~D.}{1999}]{ensemble_2}
Opitz~D., Maclin, R. (1999).
\newblock Popular ensemble methods: An empirical study.
\newblock {\em Journal of Artificial Intelligence Research}, 11:169–198.

\bibitem[\protect\astroncite{Park et~al.}{2025}]{park_forecasting_topic_trends}
Park, Y., Lim, S., Gu, C., Syafiandini, A.~F., and Song, M. (2025).
\newblock Forecasting topic trends of blockchain utilizing topic modeling and
  deep learning-based time-series prediction on different document types.
\newblock {\em Journal of Informetrics}, 19(2).

\bibitem[\protect\astroncite{Patel et~al.}{2015}]{patel_predicting_2015}
Patel, J., Shah, S., Thakkar, P., and Kotecha, K. (2015).
\newblock Predicting stock market index using fusion of machine learning
  techniques.
\newblock {\em Expert Systems with Applications}, 42(4):2162--2172.

\bibitem[\protect\astroncite{Proietti}{2011}]{proietti_direct_iterated_multistep}
Proietti, T. (2011).
\newblock Direct and iterated multistep ar methods for difference stationary
  processes.
\newblock {\em International Journal of Forecasting}.

\bibitem[\protect\astroncite{Qian and Rasheed}{2007}]{qian_stock_2007}
Qian, B. and Rasheed, K. (2007).
\newblock Stock market prediction with multiple classifiers.
\newblock {\em Applied Intelligence}, 26(1):25--33.

\bibitem[\protect\astroncite{Rahman
  et~al.}{2025}]{rahman_integrating_deep_learning_algorithms}
Rahman, M., Hasan, M.~M., Hossain, M.~A., Das, U.~K., Islam, M.~M., Karim,
  M.~R., Faiz, H., Hammad, Z., Sadiq, S., and Alam, M. (2025).
\newblock Integrating deep learning algorithms for forecasting
  evapotranspiration and assessing crop water stress in agricultural water
  management.
\newblock {\em Journal of Environmental Management}, 375.

\bibitem[\protect\astroncite{Rokach}{2010}]{ensemble_1}
Rokach, L. (2010).
\newblock Ensemble-based classifiers.
\newblock {\em Artificial Intelligence Review}, 33:1--39.

\bibitem[\protect\astroncite{S.~Siami-Namini and
  Namin}{2018}]{lstm_arima_comparison_1}
S.~Siami-Namini, N.~T. and Namin, A.~S. (2018).
\newblock {A Comparison of ARIMA and LSTM in Forecasting Time Series}.
\newblock {\em 2018 17th IEEE International Conference on Machine Learning and
  Applications}, pages 1394--1401.

\bibitem[\protect\astroncite{Sako~K}{2022}]{lstm_ml_comparison}
Sako~K, Mpinda~B.N., R.~P. (2022).
\newblock {Neural Networks for Financial Time Series Forecasting}.
\newblock {\em Entropy}, 24(5):657.

\bibitem[\protect\astroncite{Saltzman and
  Yung}{2018}]{saltzman_uncertainty_2018}
Saltzman, B. and Yung, J. (2018).
\newblock A machine learning approach to identifying different types of
  uncertainty.
\newblock {\em Economics Letters}.

\bibitem[\protect\astroncite{Schorfheide}{2005}]{schorfheide_forecasting_under_misspecification}
Schorfheide, F. (2005).
\newblock Var forecasting under misspecification.
\newblock {\em Journal of Econometrics}, 128(1):99--136.

\bibitem[\protect\astroncite{Sepp~Hochreiter}{1997}]{lstm_description}
Sepp~Hochreiter, J.~S. (1997).
\newblock {Long Short-Term Memory}.
\newblock {\em Neural Computatation}, 9(8):1735–1780.

\bibitem[\protect\astroncite{Sharma
  et~al.}{2023}]{sharma_time_series_forecasting_fb_prophet}
Sharma, K., Sharma, R., and Gansean, G. (2023).
\newblock Time series forecasting using fb-prophet: A case study on stock
  market prediction.
\newblock In {\em Proceedings of the 2nd International Conference on Data
  Science and Applications (ICDSA 2023)}, volume 3445 of {\em CEUR Workshop
  Proceedings}, page 59–65.

\bibitem[\protect\astroncite{Shi
  et~al.}{2024}]{shi_cnn_lstm_carbon_price_forecasting}
Shi, H., Wei, A., Xu, X., Zhu, Y., Hu, H., and Tang, S. (2024).
\newblock A cnn-lstm based deep learning model with high accuracy and
  robustness for carbon price forecasting: A case of shenzhen's carbon market
  in china.
\newblock {\em Journal of Environmental Management}, 352.

\bibitem[\protect\astroncite{Stock and Watson}{1998}]{ensemble_4}
Stock, J.~H. and Watson, M.~W. (1998).
\newblock {A Comparison of Linear and Nonlinear Univariate Models for
  Forecasting Macroeconomic Time Series}.
\newblock Technical Report 6607.

\bibitem[\protect\astroncite{Sushmi and
  Subbulekshmi}{2025}]{sushmi_real_time_irradiance_forecasting}
Sushmi, N. and Subbulekshmi, D. (2025).
\newblock Real-time ultra short-term irradiance forecasting using a novel r-gru
  model for optimizing pv controller dynamics.
\newblock {\em Results in Engineering}, 26.

\bibitem[\protect\astroncite{Taieb
  et~al.}{2012}]{taieb_review_comparison_strategies}
Taieb, S., Bontempi, G., Atiya, A., and Sorjamaa, A. (2012).
\newblock A review and comparison of strategies for multi-step ahead time
  series forecasting based on the nn5 forecasting competition.
\newblock {\em Expert Systems with Applications}, 39(8):7067–7083.

\bibitem[\protect\astroncite{Taieb and
  Hyndman}{2012}]{taieb_recursive_direct_multi_step}
Taieb, S. and Hyndman, R.~J. (2012).
\newblock Recursive and direct multi-step forecasting: the best of both worlds.
\newblock Technical Report 19/12, Monash University, Department of Econometrics
  and Business Statistics.

\bibitem[\protect\astroncite{Tiao and
  Xu}{1993}]{tiao_robustness_maximum_likelihood}
Tiao, G. and Xu, D. (1993).
\newblock Robustness of maximum likelihood estimates for multi-step
  predictions: the exponential smoothing case.
\newblock {\em Biometrika}, 80:623–641.

\bibitem[\protect\astroncite{Tobback et~al.}{2016}]{tobback_belgian_2016}
Tobback, E., Naudts, H., Daelemans, W., de~Fortuny, E.~J., and Martens, D.
  (2016).
\newblock Belgian economic policy uncertainty index: Improvement through text
  mining.
\newblock {\em International Journal of Forecasting}, 34(2):355--365.

\bibitem[\protect\astroncite{Tsay}{1993}]{tsay_comment_adaptive_forecasting}
Tsay, R. (1993).
\newblock Comment: Adaptive forecasting.
\newblock {\em Journal of Business and Economic Statistics}, 11(2):140–142.

\bibitem[\protect\astroncite{Vancsura~L.}{2023}]{lstm_arima_comparison_4}
Vancsura~L., Tatay~T., B.~T. (2023).
\newblock {Evaluating the Effectiveness of Modern Forecasting Models in
  Predicting Commodity Futures Prices in Volatile Economic Times}.
\newblock {\em Risks}, 11(2):27.

\bibitem[\protect\astroncite{Weiss}{1991}]{weiss_multi_step_estimation}
Weiss, A. (1991).
\newblock Multi-step estimation and forecasting in dynamic models.
\newblock {\em Journal of Econometrics}, 48:135–149.

\bibitem[\protect\astroncite{Weiss and
  Andersen}{1984}]{weiss_andersen_estimating_time_series}
Weiss, A. and Andersen, A. (1984).
\newblock Estimating time series models using the relevant forecast evaluation
  criterion.
\newblock {\em Journal of the Royal Statistical Society A}, 147:484–487.

\bibitem[\protect\astroncite{Weng et~al.}{1999}]{shi_1999}
Weng, B., Lu, L., Wang, X., Megahed, F.~M., and Martinez, W. (1999).
\newblock Improving the accuracy of nonlinear combined forecasting using neural
  networks.
\newblock {\em Expert Systems with Applications}, 16(1):49--54.

\bibitem[\protect\astroncite{Weng et~al.}{2018}]{weng_macroeconomic_2018}
Weng, B., Lu, L., Wang, X., Megahed, F.~M., and Martinez, W. (2018).
\newblock Macroeconomic indicators alone can predict the monthly closing price
  of major u.s. indices: Insights from artificial intelligence, time-series
  analysis and hybrid models.
\newblock {\em Applied Soft Computing}.

\bibitem[\protect\astroncite{Yun
  et~al.}{2023}]{yun_interpretable_stock_price_forecasting}
Yun, K., Yoon, S., and Won, D. (2023).
\newblock Interpretable stock price forecasting model using genetic
  algorithm-machine learning regressions and best feature subset selection.
\newblock {\em Expert Systems with Applications}, 213.

\end{thebibliography}

\end{document}